\documentstyle[preprint,prd,aps,draft]{revtex}
\begin{document}
\draft
\title{$B_c$ spectroscopy in a quantum-chromodynamic potential model}
\author{Suraj N. Gupta and James M. Johnson}
\address{ Department of Physics, Wayne State University, Detroit,
	Michigan 48202}
\date{August 1, 1995}
\maketitle
\begin{abstract}

	We have investigated $B_c$ spectroscopy with the use
of a quantum-chromodynamic potential model which was recently used
by us for the light-heavy quarkonia.  We give our predictions for
the energy levels and the $E$1 transition widths.  We also find,
rather surprisingly, that although $B_c$ is not a light-heavy system,
the heavy quark effective theory with the inclusion of the
$m_b^{-1}$ and $m_b^{-1}\ln m_b$ corrections is as successful for $B_c$ as it
is for $B$ and $B_s$.

\end{abstract}
\pacs{12.39.Pn,12.39.Hg,14.40.Lb,14.40.Nd}
\narrowtext
\clearpage

\section{INTRODUCTION}

	$B_c$ spectroscopy has been investigated by several authors
\cite{kwong,chen,eichten,gershtein}
in recent years by using different models and arriving at different
predictions for this hitherto unobserved quarkonium.
Although $B_c$ consists of heavy quarks, its decay modes are not the
same as those of $b\bar{b}$ and $c\bar{c}$. Indeed, because of
flavor conservation in strong and electromagnetic interactions,
the $B_c$ ground state can only decay weakly, which makes
it particularly interesting for the study of weak interactions.

	We shall present our results for the $B_c$ spectroscopy
by using a quantum-chromodynamic potential model which was recently
used by us for the light-heavy quarkonia\cite{light}.  An essential feature
of our model is the inclusion of the one-loop radiative corrections
in the quantum-chromodynamic potential, which is known to be
responsible for the remarkable agreement between the theoretical and
experimental results for spin splittings in the $b\bar{b}$ and $c\bar{c}$
spectra\cite{bbcc}.  Another advantage of our model is that it is based
on a nonsingular form of the quarkonium potential, and thus avoids
the use of an illegitimate perturbative treatment.

	The choice of potential parameters for $B_c$ in the absence of
experimental data will be discussed in Sec.~II, while its spectrum and $E$1
transition widths will be given in Sec.~III.  We shall also demonstrate
the rather surprising result that although $B_c$ is not a light-heavy
system, the heavy quark effective theory \cite{hqet} with the inclusion of the
$m_b^{-1}$ and $m_b^{-1}\ln m_b$ corrections is as successful for $B_c$ as it
is for $B$ and $B_s$.

\section{$B_c$ Potential Parameters}

	Our model is based on the Hamiltonian
\begin{equation}
H=H_0+V_p+V_c,
\end{equation}
where
\begin{equation}
H_0=(m_c^2+{\bf p}^2)^{1/2}+(m_b^2+{\bf p}^2)^{1/2}
\end{equation}
is the relativistic kinetic energy, and $V_p$ and $V_c$ are
nonsingular quasistatic perturbative and confining potentials,
which are fully given in Ref.~5.  The perturbative
potential with the one-loop corrections involves the parameters
$m_c$, $m_b$, $\mu$, and $\alpha_s$, while the phenomenological
scalar-vector exchange confining potential involves, besides the quark
masses, the parameters $A$ and $B$ as well as an additive constant $C$.

	We expect the dynamics of $B_c$ to be largely dependent on
the lighter quark $c$.  Therefore, in the absence of experimental data,
we assume that $m_c$, $\mu$, $\alpha_s$, $A$ and $B$ for $B_c$
have the same values as those for $c\bar{c}$, while $m_b$ for
$B_c$ is obtainable from its value for $b\bar{b}$ by the QCD
transformation relation.  The constant $C$ is usually fixed
by the experimental value of the quarkonium ground state, but
here we make the {\it ad hoc} assumption that $C$ is equal to the
average of its values for $c\bar{c}$ and $b\bar{b}$, so that
\begin{equation}
C_{b\bar{c}}={\textstyle\frac{1}{2}}\left(C_{c\bar{c}}+C_{b\bar{b}}\right).
\label{adhoc}
\end{equation}

	We give in Tables~I and~II the spectra and parameter values for
$c\bar{c}$ and $b\bar{b}$ by updating our earlier results \cite{bbcc}
with the use of the latest experimental
data provided by the Particle Data Group \cite{pdg}.  The values
of $\alpha_s$ for $c\bar{c}$ and $b\bar{b}$ in these tables approximately
satisfy the QCD transformation relation
\begin{equation}
\alpha_s^\prime=\frac{\alpha_s}{1+\beta_0(\alpha_s/4\pi)
	\ln ({\mu^\prime}^2/\mu^2)}\ ,
\end{equation}
where $\beta_0=11-\frac{2}{3}n_f$, $n_f=3$.
We also note that, according to the QCD transformation relation
\begin{equation}
m^\prime=m\left(\frac{\alpha_s^\prime}{\alpha_s}
	\right)^{2\gamma_0/\beta_0},
\end{equation}
with $\gamma_0=2$, the value of $m_b$ in Table~II for $\mu=\mu_{b\bar{b}}$
leads to
\begin{equation}
m_b=5.453\ {\rm GeV\quad for}\quad\mu=\mu_{c\bar{c}}.
\end{equation}

\section{$B_c$ Spectra and $E$1 Transitions}

	We have calculated the $B_c$ spectrum by using the potential
parameters in Sec.~II and following the same procedure as was applied to the
light-heavy quarkonia in Ref.~5.  The theoretical results for the
energy levels, together with the ${}^3P_1'$-${}^1P_1'$ mixing angle
arising from the spin-orbit mixing terms, are given in
Table~III.  In this table, one set of results corresponds to the direct
use of our model, while the other two sets are obtained
by means of heavy quark expansions of our potentials with the
inclusion of the $m_b^{-1}$ and $m_b^{-1}\ln m_b$ corrections as well as
without these corrections.
Our results numerically differ to varying degrees from those of Chen
and Kuang\cite{chen}, Eichten and Quigg\cite{eichten}, and Gershtein
{\it et al.}\cite{gershtein}, and a comparison of various results
for the lowest $S$ states is shown in Table~IV.

	It should be noted that only the energy differences among the
energy levels are predicted by our potential model, while the absolute
energy levels have been obtained by making use of the
assumption~(\ref{adhoc}).
A variation of the parameter $C_{b\bar{c}}$ will cause a common
shift of our energy levels in Tables~III and~IV.

	In Table~V, we give the results for the $E1$ transition widths for $B_c$
by using the formulae
\begin{eqnarray}
\Gamma_{E1}({}^3S_1 \rightarrow {}^3P_J) &=& \frac{4}{9}
        \frac{2J+1}{3}\alpha \langle e_Q\rangle^2 k_J^3
	|r_{fi}|^2,\nonumber \\
\Gamma_{E1}({}^3P_J \rightarrow {}^3S_1) &=&\frac{4}{9}\alpha
	\langle e_Q\rangle^2 k_J^3 |r_{fi}|^2,\\
\Gamma_{E1}({}^1P_1 \rightarrow {}^1S_0) &=&\frac{4}{9}\alpha
	\langle e_Q\rangle^2  k_J^3 |r_{fi}|^2,\nonumber\\
\Gamma_{E1}({}^1S_0 \rightarrow {}^1P_1) &=&\frac{4}{3}\alpha
	\langle e_Q\rangle^2  k_J^3 |r_{fi}|^2,\nonumber
\end{eqnarray}
where the mean charge $\langle e_Q\rangle$ is given by \cite{eichten}
\begin{equation}
\langle e_Q\rangle=\frac{m_be_c-m_ce_{\bar{b}}}{m_b+m_c}.
\end{equation}
The photon energies for the $E1$ transition widths have been obtained from
the energy difference of the initial and final $b\bar{c}$ states by
taking into account the recoil correction.

	Apart from numerical differences, our results in Table~V differ
from those of Ref.~3 in two respects.  In Ref.~3, the results
for $r_{fi}$ are the same for all $1P\rightarrow 1S$ transitions
as well as for all $2S\rightarrow 1P$ transitions\cite{pnote}.
We have a different value for $r_{fi}$ for each transition because
our nonsingular potential allows us to include the spin-dependent
terms in the unperturbed Hamiltonian.  Furthermore, in Ref.~3
some of the widths for transitions involving the mixed $P$ states
are vanishingly small, while this is not the case in our treatment.
This difference indicates that our potential gives rise to a larger
spin-orbit mixing effect.

	Finally, a comparison of our results for $B_c$ in Table~III
with the corresponding results for $B$ and $B_s$ in Ref.~5
shows that the heavy quark expansion with the $m_b^{-1}$ and $m_b^{-1}\ln m_b$
corrections is as successful for $B_c$ as it is for $B$ and $B_s$.  This is
rather surprising because the heavy quark effective theory has been
generally applied to the light-heavy quarkonia.

\acknowledgments
        This work was supported in part by the U.S. Department of Energy
under Grant No.~DE-FG02-85ER40209.

\narrowtext

\mediumtext
\begin{table}
\caption{\label{cc} $c\bar{c}$ spectrum and parameter values.
The energy levels are given in MeV.}
\begin{tabular}{ldr@{$\pm$}l}
\hspace*{1pt}&Theory&\multicolumn{2}{c}{Expt.}\\
\tableline
$1\;{}^1S_0\, (\eta_c)$&	2979.1&		$2978.8$&$1.9$ \\
$1\;{}^3S_1\, (J/\psi)$&	3096.9&		$3096.88$&$0.04$ \\
$2\;{}^1S_0\, (\eta_c')$&	3617.9&		\multicolumn{2}{c}{} \\
$2\;{}^3S_1\, (\psi')$&		3685.9&		$3686.00$&$0.09$ \\
$1\;{}^3P_0\, (\chi_{c0})$&  	3415.2&		$3415.1$&$1$ \\
$1\;{}^3P_1\, (\chi_{c1})$&  	3510.8&		$3510.53$&$0.12$ \\
$1\;{}^3P_2\, (\chi_{c2})$&  	3556.5&		$3556.17$&$0.13$ \\
$1\;{}^1P_1\, (h_c)$&		3526.4&		$3526.14$&$0.24$\\
$m_c$ (GeV)&			2.212\\
$\mu_{c\bar{c}}$ (GeV)&		2.942\\
$\alpha_s$&			0.306\\
$A$ (GeV$^2$)&			0.181\\
$B$&				0.244\\
\end{tabular}
\end{table}

\begin{table}
\caption{\label{bb} $b\bar{b}$ spectrum and parameter values.
The energy levels are given in MeV.}
\begin{tabular}{ldr@{$\pm$}l}
\hspace*{1pt}&Theory&\multicolumn{2}{c}{Expt.}\\
\tableline
$1\;{}^1S_0\, (\eta_b)$&	9407.6&		\multicolumn{2}{c}{} \\
$1\;{}^3S_1\, (\Upsilon)$&	9460.3&		$9460.37$&$0.21$ \\
$2\;{}^1S_0\, (\eta_b')$&	9990.5&		\multicolumn{2}{c}{} \\
$2\;{}^3S_1\, (\Upsilon')$&	10016.1&	$10023.30$&$0.31$ \\
$3\;{}^1S_0\, (\eta_b'')$&	10338.0&	\multicolumn{2}{c}{} \\
$3\;{}^3S_1\, (\Upsilon'')$&	10357.9&	$10355.3$&$0.5$ \\
$1\;{}^3P_0\, (\chi_{b0})$&  	9861.9&		$9859.8$&$1.3$ \\
$1\;{}^3P_1\, (\chi_{b1})$&  	9893.4&		$9891.9$&$0.7$ \\
$1\;{}^3P_2\, (\chi_{b2})$&  	9914.2&		$9913.2$&$0.6$ \\
$1\;{}^1P_1\, (h_b)$&		9900.8&		\multicolumn{2}{c}{} \\
$2\;{}^3P_0\, (\chi_{b0}')$&  	10228.8&	$10232.1$&$0.6$ \\
$2\;{}^3P_1\, (\chi_{b1}')$&  	10253.5&	$10255.2$&$0.5$ \\
$2\;{}^3P_2\, (\chi_{b2}')$&  	10269.8&	$10268.5$&$0.4$ \\
$2\;{}^1P_1\, (h_b')$&		10259.4&	\multicolumn{2}{c}{} \\
$m_b$ (GeV)&			5.406\\
$\mu_{b\bar{b}}$ (GeV)&		3.435\\
$\alpha_s$&			0.283\\
$A$ (GeV$^2$)&			0.184\\
$B$&				0.388\\
\end{tabular}
\end{table}

\begin{table}
\caption{\label{bc} $B_c$ energy levels
in MeV.  Effective theory results are given with the $m_b^{-1}$ and
$m_b^{-1}\ln m_b$ corrections as well as in the limit of
$m_b\rightarrow\infty$. }
\bigskip
\begin{tabular}{lddd}
\hspace*{1pt}&		Theory&Effective theory&$m_b\rightarrow\infty$\\
\tableline
$1\;{}^1S_0\,\ (B_c)$&		6246.9& 6246.9& 6246.9\\
$1\;{}^3S_1\,\ (B_c^\star)$&	6308.0& 6311.0& 6246.9\\
$2\;{}^1S_0\,$&			6852.8& 6853.5& 6828.6\\
$2\;{}^3S_1\,$&			6885.9& 6887.9& 6828.6\\
$1\;{}^3P_0\,$&			6688.6& 6693.8& 6716.8\\
$1\;{}^3P^\prime_1\,$&		6737.5& 6737.9& 6716.8\\
$1\;{}^1P^\prime_1\,$&		6757.3& 6758.1& 6752.3\\
$1\;{}^3P_2\,$&			6773.2& 6772.3& 6752.3\\
$\theta$&		25.6$^\circ$&  28.8$^\circ$&  35.6$^\circ$
\end{tabular}
\end{table}

\begin{table}
\caption{$B_c$ and $B_c^\star$ energy levels in MeV in our model and some
earlier
potential models. }
\bigskip
\begin{tabular}{lcccc}
	&G-J&	Chen-Kuang&	Eichten-Quigg&	Gershtein {\it et al.}\\
\tableline
${}^1S_0\ (B_c)$& 	6,247&	6,310&	6,264&	6,253 \\
${}^3S_1\ (B_c^\star)$&	6,308&	6,355&	6,337& 	6,317\\
$B_c^\star -B_c$&	61&	45&	73&	64
\end{tabular}
\end{table}

\begin{table}
\caption{\label{E1} $E1$ transition widths for $B_c$. }
\bigskip
\begin{tabular}{lddd}
Transition&  Photon energy (MeV)&  $|r_{fi}|$ (GeV$^{-1}$)&  $\Gamma_{E1}$
(keV)\\
\tableline
$1\;{}^3P_2\rightarrow 1\;{}^3S_1$& 	449&	1.30&	73.6 \\
$1\;{}^3P_1'\rightarrow 1\;{}^3S_1$&	416&	1.19&   49.0 \\
$1\;{}^3P_1'\rightarrow 1\;{}^1S_0$&	473&	0.57&   16.6 \\
$1\;{}^3P_0\rightarrow 1\;{}^3S_1$&	370&	1.33&   43.0 \\
$1\;{}^1P_1'\rightarrow 1\;{}^1S_0$&	491&	1.08&	66.6 \\
$1\;{}^1P_1'\rightarrow 1\;{}^3S_1$&	434&	0.52&	10.5 \\
$2\;{}^3S_1\rightarrow 1\;{}^3P_2$& 	112&	1.91&	4.0 \\
$2\;{}^3S_1\rightarrow 1\;{}^3P_1'$&	147&	1.56&   3.6 \\
$2\;{}^3S_1\rightarrow 1\;{}^1P_1'$&	127&	0.80&   0.6 \\
$2\;{}^3S_1\rightarrow 1\;{}^3P_0$&	194&	1.49&   2.6 \\
$2\;{}^3S_0\rightarrow 1\;{}^1P_1'$&	95&	1.73&   3.6 \\
$2\;{}^3S_0\rightarrow 1\;{}^3P_1'$&	114&	0.78&   1.3
\end{tabular}
\end{table}


\begin{thebibliography}{99}
\bibitem{pnote} We have denoted the lowest $P$ states of quarkonia as $1P$
	in accordance with the notation followed by the Particle Data
	Group\cite{pdg}.  This notation differs from that used in Ref.~3.
\end{thebibliography}
\end{document}